\documentclass[prc,aps,amsfonts,amsmath,superscriptaddress,showpacs,floatfix,twocolumn]{revtex4}
\usepackage{graphicx} 
\usepackage{epsfig} 
\usepackage{rotating}
\usepackage{mwe}    
\usepackage{subfig}
\begin{document}

\title{Density Functional Theory studies of cluster states in nuclei}

\author{J.-P. Ebran}
\affiliation{CEA,DAM,DIF, F-91297 Arpajon, France}
\author{E. Khan}
\affiliation{Institut de Physique Nucl\'eaire, Universit\'e Paris-Sud, IN2P3-CNRS, 
F-91406 Orsay Cedex, France}
\author{T. Nik\v si\' c}
\author{D. Vretenar}
\affiliation{Physics Department, Faculty of Science, University of
Zagreb, 10000 Zagreb, Croatia}

\begin{abstract}
The framework of nuclear energy density functionals is applied to a
study of the formation and evolution of cluster states in nuclei.  The
relativistic functional DD-ME2 is used in triaxial and 
reflection-asymmetric relativistic Hartree-Bogoliubov calculations
of relatively light $N = Z$ and neutron-rich nuclei.  The role of
deformation and degeneracy of single-nucleon states in the formation
of clusters is analyzed, and interesting cluster structures are
predicted in excited configurations of Be, C, O, Ne, Mg, Si, S, Ar and
Ca $N = Z$ nuclei. Cluster phenomena in neutron-rich nuclei are 
discussed, and it is shown that in neutron-rich Be and C nuclei
cluster states occur that are characterized by molecular bonding of
$\alpha$-particles by the excess neutrons.
\end{abstract}


\date{\today}

\maketitle

\section{Introduction}

Nuclear Energy Density Functional (EDF) provides a comprehensive and accurate description of ground-state properties 
and collective excitations over the whole nuclide chart. In the last decade EDFs have also 
been successfully applied to studies of clustering phenomena, and this framework enables a consistent microscopic 
analysis of the formation and evolution of cluster structures that is not limited to the lightest nuclei 
\cite{ich11,ich12,gir13,nat,ebr13,ebr14,mar06,aru,yao14}. To describe the phenomenon of nuclear clustering already in the 
most basic EDF implementation, the self-consistent mean-field level, it is necessary to break as many spatial symmetries 
of the nuclear system as possible, and this implies a considerable computational cost. This explains the rather recent application 
of EDF-based methods to detailed quantitative studies of nuclear clustering. Consequently this approach provides a basis 
for the theoretical study of coexistence of cluster states and mean-field-type states. Cluster structures can, in fact, be considered 
as a transitional phase between the quantum liquid (nucleonic matter) phase and a crystal phase that does not 
occur in finite nuclei. Similar phase transition between the liquid and crystal phases are found in 
studies of mesoscopic systems such as quantum dots \cite{wain},
or bosons in a rotating trap \cite{yan07}.

The solid (crystal) vs. quantum liquid nature of nuclear matter has been
analyzed using the quantality parameter \cite{mot96}, defined as the
ratio of the zero-point kinetic energy of the confined nucleon to its
potential energy. The typical value obtained for nuclear matter is
characteristic for a quantum liquid phase and reflects the well-known
fact, recently confirmed by microscopic self-consistent Green's
function calculation \cite{soma}, that a nucleon in nuclear matter has
a large mean-free path of 4 to 5 fm. The quantality parameter,
however, is defined for infinite homogeneous systems and its
applicability to finite nuclei is limited by the fact that it does not
contain any nuclear mass or size dependence. Cluster states in finite
nuclei introduce an additional phase of nucleonic matter, and to
analyze localization and the phenomenon of clustering a quantity must
be considered that is sensitive to the nucleon number and size of the
nucleus. This is the localization parameter introduced in
Refs.~\cite{nat,ebr13,ebr14}. Its value increases with mass and
describes the gradual transition from a hybrid phase in light nuclei,
characterized by the spatial localization of individual nucleon states
that leads to the formation of cluster structures, toward the Fermi
liquid phase in heavier nuclei. The relationship between the
quantality and the localization parameters is detailed in Appendix A.

In this work we apply nuclear EDF to a study of the formation and evolution of cluster states in nuclei. 
The framework of nuclear EDFs and the role of spatial localization of the individual 
single-nucleon states is reviewed in section II. Section III presents an analysis of the role of deformation 
and pronounced level degeneracy on the formation of clusters, and includes a number of examples of 
cluster structures in excited states. Cluster phenomena and molecular bond in neutron-rich nuclei 
are discussed in section IV, and section V contains a short summary and conclusion of the present study.

\section{Nuclear Energy Density Functionals}

The framework of EDFs provides a 
global approach to nuclear structure and enables an accurate
description of ground-state properties and collective excitations over
the whole chart of nuclides. At a moderate computational cost modern
non-relativistic and relativistic EDFs can describe the evolution of
structure phenomena from relatively light systems to superheavy
nuclei, and from the valley of $\beta$-stability to the particle
drip-lines. 
 \begin{figure}[h!]
   
      {\includegraphics[width=0.55\textwidth]{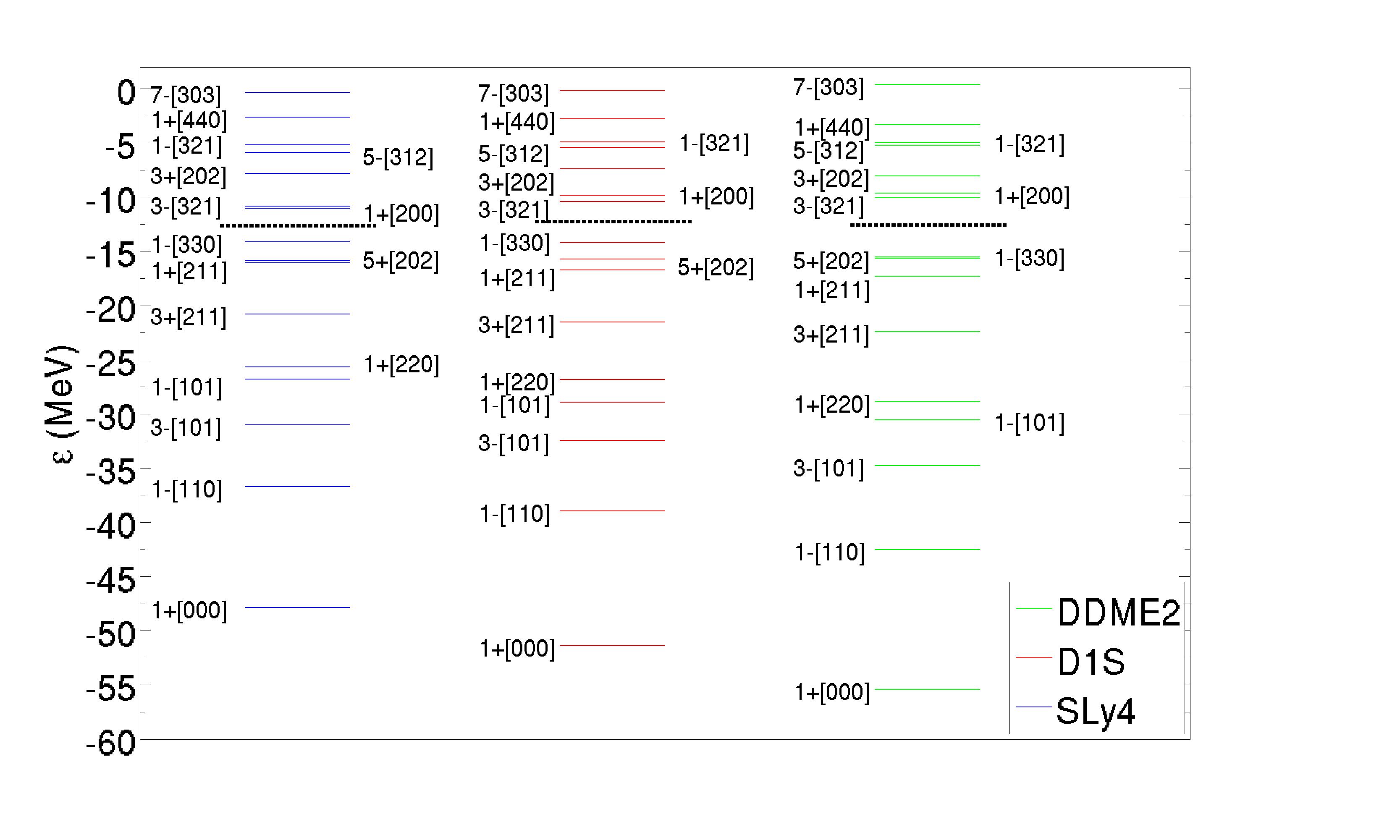}
    }
    \caption{(color online) Neutron single-particle levels of $^{36}$Ar that correspond to the SCMF solutions 
    calculated with the Skyrme functional SLy4, the Gogny effective 
interaction D1S, and the relativistic density functional DD-ME2.  
The levels are labelled by the Nilsson quantum numbers, and dotted lines 
denote the position of the Fermi level.}
    \label{fig:spectre_n}
  \end{figure}

The nuclear EDF is built from powers and gradients of ground-state nucleon densities and currents, representing distributions 
of matter, spins, momentum and kinetic energy. In principle a nuclear EDF can incorporate all short-range correlations related 
to the repulsive core of the inter-nucleon interaction, and long-range correlations mediated by nuclear resonance 
modes. An additional functional of the pairing density is included to account for effects of superfluidity in 
open-shell nuclei.

The ground-state energy and density of a given system can be determined by minimizing an EDF with respect to the
$3$-dimensional density. The self-consistent scheme introduces a local effective single-particle potential, 
such that the exact ground-state density of the interacting system of particles equals the ground-state density of the
auxiliary non-interacting system, expressed in terms of the lowest occupied single-particle orbitals. The many-body dynamics 
is represented by independent nucleons moving in local self-consistent
mean-field (SCMF) potentials that correspond to the actual density and current distributions of a given nucleus. 

\begin{figure}[ht]
    \subfloat[SLy4\label{fig:Ar36_SLy4_EvsQ}]{%
      \includegraphics[width=0.46\textwidth]{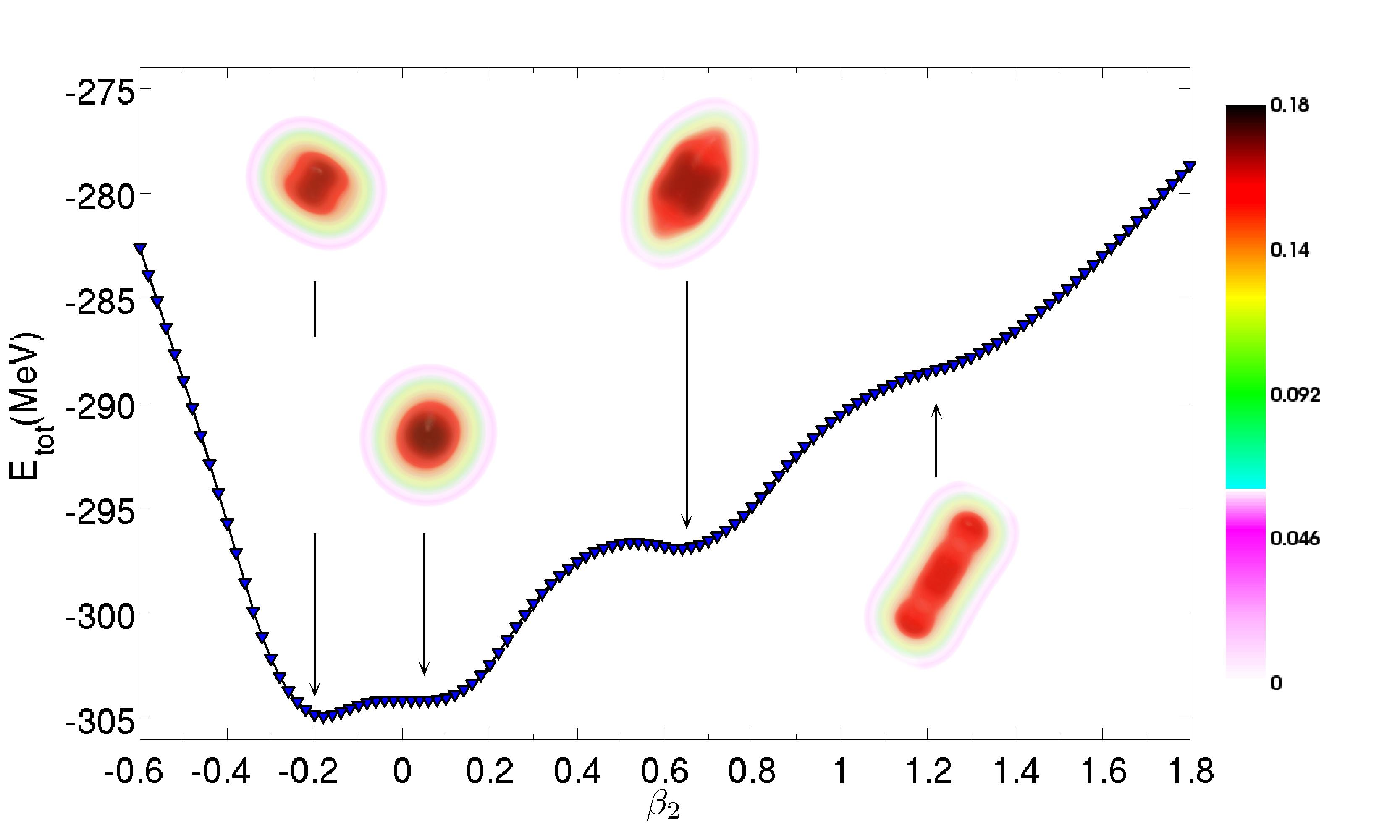}
    }\\
    \subfloat[D1S\label{fig:Ar36_D1S_EvsQ}]{%
      \includegraphics[width=0.46\textwidth]{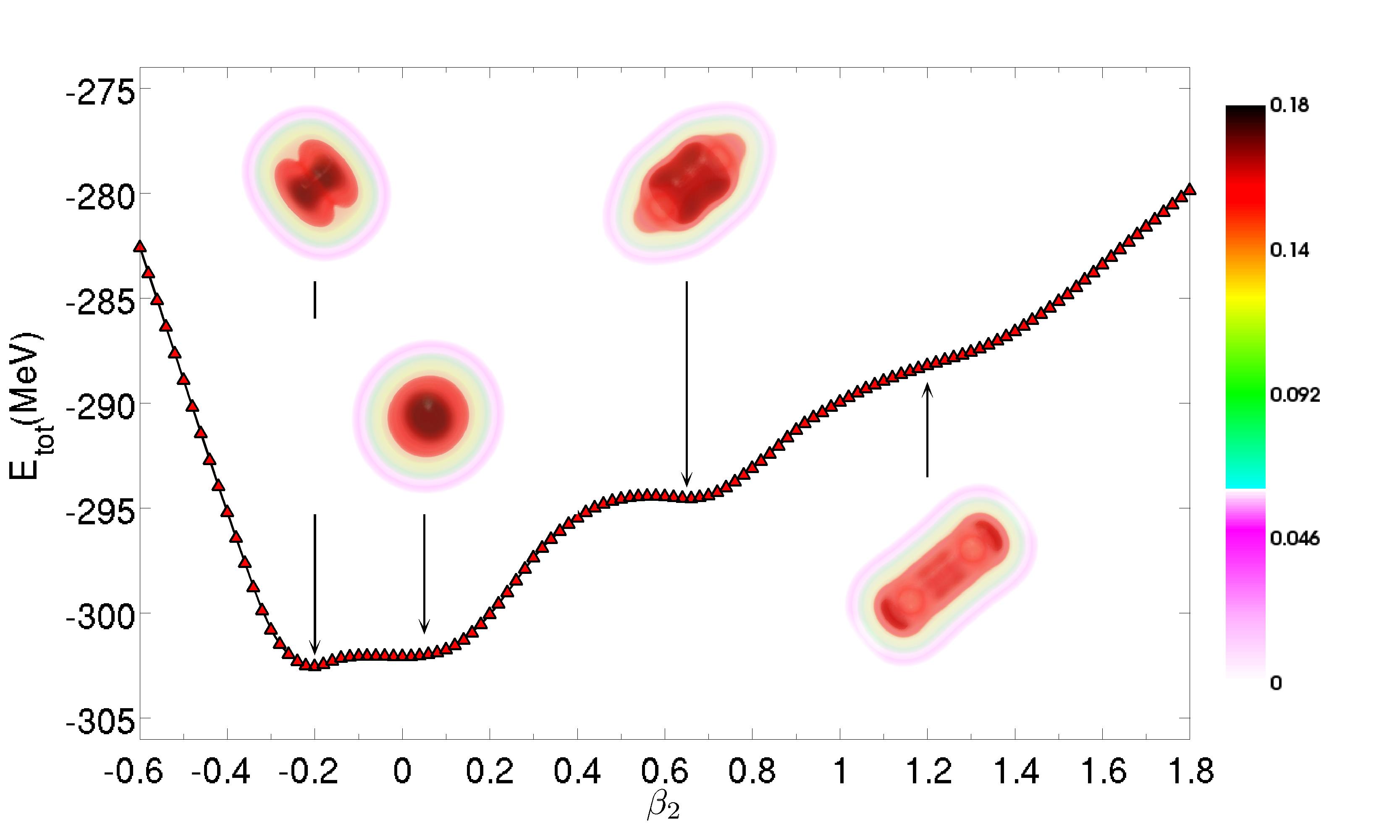}
    }\\
    \subfloat[DD-ME2\label{fig:Ar36_DDME2_EvsQ}]{%
      \includegraphics[width=0.46\textwidth]{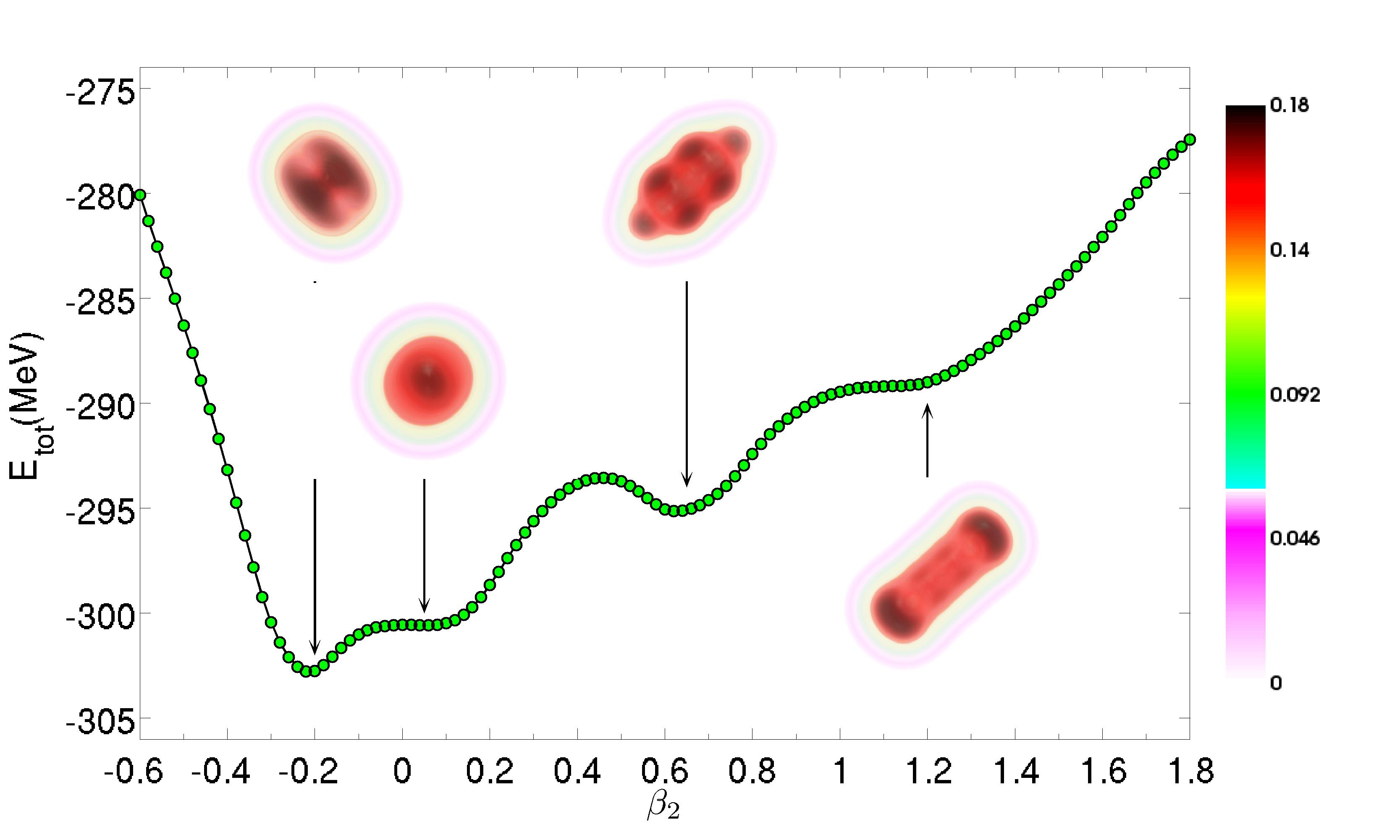}
    }
    \caption{(color online) Self-consistent binding energy curves of $^{36}$Ar as 
    functions of the quadrupole deformation parameter $\beta_2$, calculated 
    with the functionals SLy4, D1S and DD-ME2. The insets display the 
    corresponding intrinsic nucleon density distributions in the reference frame defined
	by the principal axes of the nucleus.}
    \label{fig:Ar36_EvsQ}
  \end{figure}

A broad range of nuclear structure phenomena have been analyzed using 
Skyrme, Gogny and relativistic EDFs \cite{Ben03,Lal04,Sto07aR,Erl11,Vre05,Men06,Nik11}. 
These global functionals present different realizations of a universal nuclear EDF 
governed by the underlying theory of strong interactions. With relatively small 
sets of global parameters determined 
by empirical properties of nucleonic matter and data on finite nuclei, structure 
models based on Skyrme, Gogny or relativistic functionals provide a consistent 
description of a vast quantity of nuclear data. 
Even though results for ground-state observables (e.g. binding energies, charge 
radii) obtained with different functionals are rather similar and of comparable agreement 
with data, calculated quantities that are not directly observable can show marked 
differences. One such quantity is the auxiliary local SCMF 
potential. In Fig.~\ref{fig:spectre_n} we plot the neutron single-particle levels 
of $^{36}$Ar calculated with the Skyrme functional SLy4 \cite{SLy4}, the Gogny effective 
interaction D1S \cite{BGG.84,BGG.91}, and the relativistic density functional DD-ME2 \cite{DD-ME2}. 
The levels are labelled by the Nilsson quantum numbers, and correspond to ground-state 
SCMF solutions with the assumption of an axially symmetric quadrupole deformation. Dotted lines 
denote the position of the Fermi level. Even though all three functionals predict very similar 
ground-state properties (cf. also Fig.~\ref{fig:Ar36_EvsQ}) and, therefore, similar ordering and 
density of levels close to the Fermi surface, the depths of the corresponding SCMF 
potentials are markedly different. The deepest potential correspond to the relativistic functional 
DD-ME2 ($-82.4$ MeV), whereas the potential of the Skyrme functional SLy4 is fairly shallow ($-72.4$ MeV). 
The position of the 1s state shows that the effective depth of the D1S potential lies between
the ones of DD-ME2 and SLy4. One finds the same picture for the proton 
states except, of course, for the effect of Coulomb repulsion. 

In Ref.~\cite{nat} we found 
qualitatively the same difference for the SCMF potentials of $^{20}$Ne calculated with 
SLy4 and DD-ME2. Even though the SCMF potential is not an observable, a deeper 
confining potential leads to a more pronounced localization of the single nucleon wave 
functions and enhances the probability of formation cluster structures in excited states close to 
the energy threshold for $\alpha$-particle emission. The formation of nuclear clusters is 
similar to a transition from a superfluid to a Mott insulator phase in a 
gas of ultracold atoms held in a three-dimensional optical lattice potential 
\cite{gre02,Jaksch98}. As the potential depth of the lattice is increased, a transition 
is observed from a phase in which each atom is spread out over the entire 
lattice, to the insulating phase in which atoms are localized with no 
phase coherence across the lattice. In the nuclear case one cannot, of course, 
vary the depth of the single-nucleon potential because the nucleus is a self-bound system. 
However, the same effect can be obtained by performing self-consistent calculations 
using different EDFs or effective interactions, as illustrated 
in Fig.~\ref{fig:spectre_n} for SLy4, D1S and DD-ME2.

To investigate the role of deformation in the formation of clusters,
we perform deformation-constrained SCMF calculations by imposing
constraints on the mass multipole moments of a nucleus. The
corresponding equations (Schr\"odinger-like for non-relativistic
functionals, or Dirac-like for relativistic EDFs, with the Hamiltonian
defined as the functional derivative of the corresponding EDF with
respect to density) are solved in the intrinsic frame of reference
attached to the nucleus, in which the shape of the nucleus can be
arbitrarily deformed. In the present study we employ SCMF models that
allow breaking both the axial and reflection symmetries \cite{Ben03}.
As an illustration in Fig.~\ref{fig:Ar36_EvsQ} we display the binding
energies of the self-conjugate nucleus $^{36}$Ar as functions of the
axial quadrupole deformation parameter $\beta_2$, calculated with SLy4
and D1S using the Hartree-Fock-Bogoliubov (HFB) model \cite{sto05,BGG.91},
and with the functional DD-ME2 employing the relativistic
Hartree-Bogoliubov (RHB) approach \cite{Vre05}. Pairing correlations
are taken into account by a delta-pairing force for calculations with
the Skyrme functional, whereas for the RHB calculations with DD-ME2
the pairing interaction is separable in momentum space, and determined
by two parameters adjusted to reproduce the Gogny pairing gap in
symmetric nuclear matter \cite{go96}. The curves of the total energy as a function
of quadrupole deformation are obtained in a SCMF approach by imposing
a constraint on the axial quadrupole moment. The parameter $\beta_2$
is directly proportional to the intrinsic mass quadrupole moment.  For
all three functionals the calculated equilibrium shape of $^{36}$Ar is
a slightly oblate, axially symmetric quadrupole ellipsoid with 
$\beta_2 \approx -0.2$. For the equilibrium deformation and few
additional values of $\beta_2$, in the insets of
Fig.~\ref{fig:Ar36_EvsQ} we also include the corresponding intrinsic
nucleon density distributions in the reference frame defined by the
principal axes of the nucleus. Here one already observes an
interesting effect that was previously noted in our studies of
Refs.~\cite{nat,ebr13,ebr14}, namely that deeper potentials lead to a
more pronounced spatial localization of nucleonic densities. In
general, we find that relativistic functionals, when compared to 
Skyrme and Gogny functionals, are characterized by deeper SCMF
potentials.  As noted in Ref.~\cite{nat}, the depth of a relativistic
potential is determined by the difference between two large fields: an
attractive (negative) Lorentz scalar potential of magnitude around 400
MeV, and a repulsive Lorentz vector potential of roughly 320 MeV (plus
the repulsive Coulomb potential for protons). The sum of these
potentials (about 700 MeV) determines the effective single-nucleon
spin-orbit force in a unique way, whereas in a non-relativistic EDF
framework the spin-orbit potential is included in a purely
phenomenological way, with a strength parameter adjusted to empirical
energy spacings between spin-orbit partner states. In the relativistic
case the scalar and vector fields determine both the effective
spin-orbit force and the SCMF potential, and the latter is generally
found to be deeper than the non-relativistic mean-field potentials. In
the following sections we, therefore, perform SCMF calculations based
on the relativistic functional DD-ME2, which predicts equilibrium
density distributions that are more localized, often with pronounced
cluster structures.

\section{Deformations and excited configurations}

A unique feature of light nuclei is the coexistence of the nuclear mean-field 
and cluster structures, as expressed 
by the well-known Ikeda diagram \cite{Ike68,4,5,hor11,Hor12}. A certain degree of 
localization of nucleonic densities is already present in mean-field ground-state 
configurations \cite{20,nat,schu12}, and this facilitates the formation of cluster 
structures in excited states. Close to the particle emission threshold continuum 
effects become important for a quantitative description of nuclear clustering \cite{oko12}. Deformation in 
light nuclei plays, of course, an important role in the formation of clusters \cite{21,
5,ebr14,ich11,ich12}. The relationship between $\alpha$-clusters and single-particle 
states in deformed nuclei has been extensively studied \cite{5,hor11,abe94}.
For instance, the Bayman-Bohr theorem \cite{bay58} states that the 
SU(3) shell model wave function of a ground state is in most cases 
equivalent to the cluster Brink wave function in the limit when the inter-alpha 
distance vanishes. However, this important link only relates a cluster wave function to a mean-field 
type one in this specific limit. The present EDF-based 
approach allows one to go a step further and establish a link between cluster
states and the single-particle spectrum.

\begin{figure}[h!]
    \subfloat[$^{12}$C\label{fig:C12_Emoychain}]{%
      \includegraphics[width=0.47\textwidth]{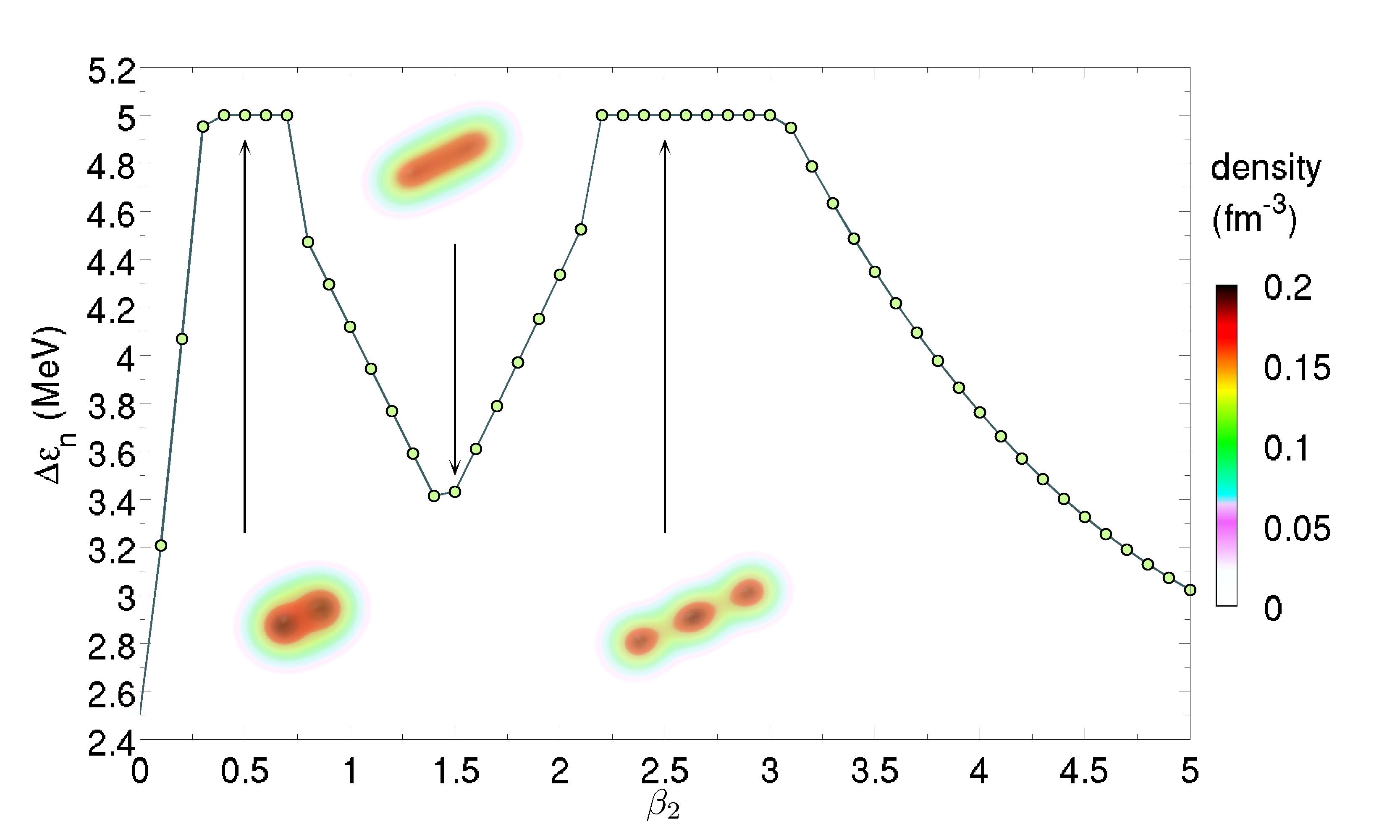}
    }
    \hfill
    \subfloat[$^{20}$Ne\label{fig:Ne20_Emoychain}]{%
      \includegraphics[width=0.47\textwidth]{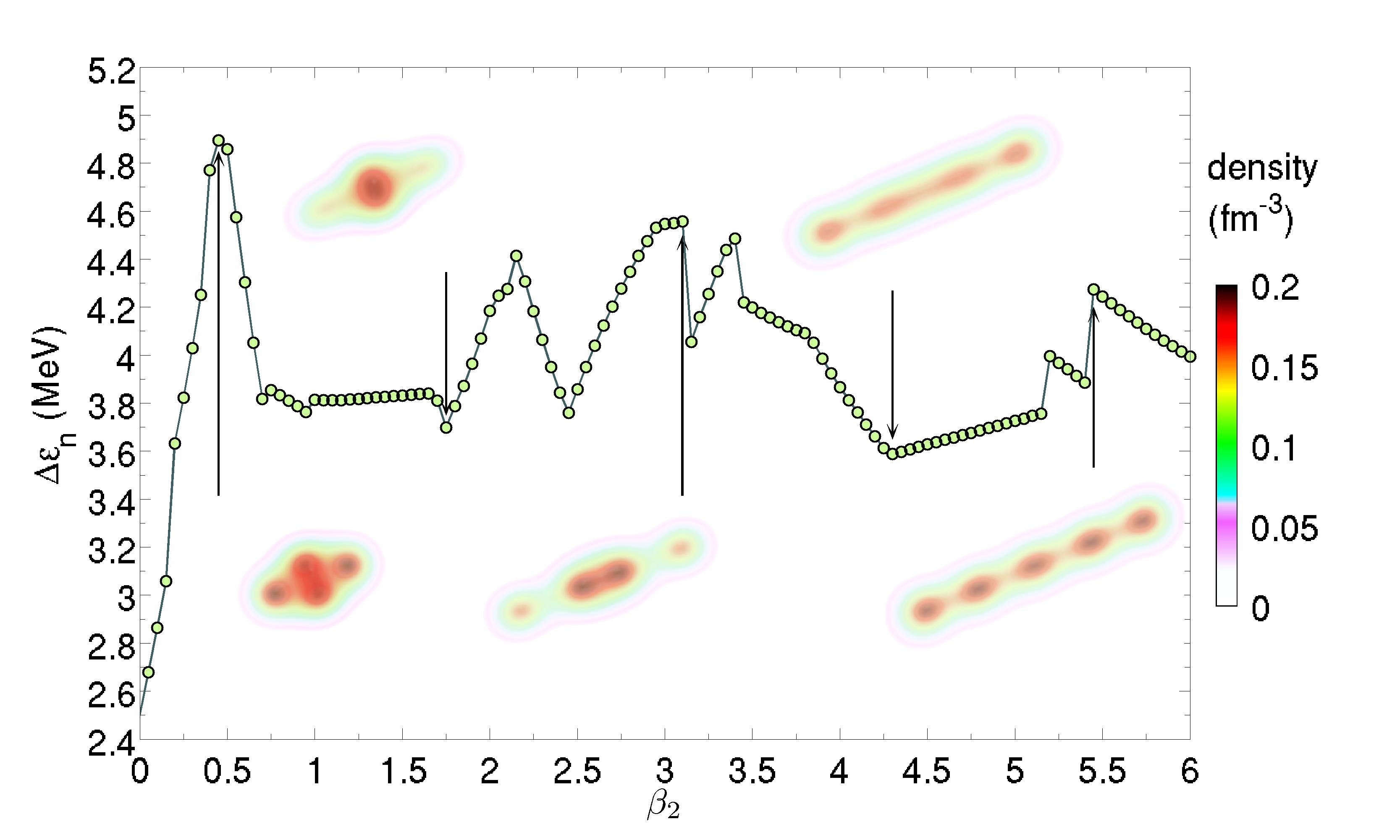}
    }
    \caption{(color online) Mean value of the energy gap between consecutive occupied neutron levels 
    as a function of the axial quadrupole deformation parameter $\beta_2$ for $^{12}$C (a) and 
    $^{20}$Ne (b). The insets display the total nucleonic density at the
    corresponding deformation. To limit the vertical scale the maximum mean value of the energy gap 
    in the plot does not exceed 5 MeV.}
    \label{fig:Emoychain}
  \end{figure}

\subsection{Axially-symmetric quadrupole deformations}

As stated by Rae \cite{rae89}, the degeneracy of single-nucleon
states at a given deformation could generate clusters because of
levels crossing. Here we analyze how degeneracy affects the formation
of $\alpha$ clusters in self-conjugate nuclei.  As noted by Aberg
\cite{abe94}, an isolated level of the single-particle energy spectrum
in a deformed self-conjugate $N=Z$ nucleus can correspond to an
alpha-cluster, because of both time-invariance symmetry and isospin
symmetry: two protons and two neutrons have similar wave functions,
and the localization of these functions facilitates the formation of
$\alpha$-clusters. Hence, pronounced level degeneracy (or isolated
levels in the case of alpha-clustering) allows to explain: i) why
$N=Z$ and deformed nuclei favor cluster formation, ii) the link
between the depth of the confining potential and cluster formation
and, iii) why cluster structures mainly occur in light nuclei. The
second point is related to the fact that pronounced degeneracy is
driven by the depth of the potential \cite{Rin80}, and this issue has
already been analyzed in our previous studies \cite{nat,ebr13,ebr14}.
The answer to the third question comes from the fact that level
density is generally smaller in lighter nuclei and this favors the occurrence 
of isolated single-particle levels or degeneracy
at certain deformations. 
\begin{figure}[h!]
\hspace{-1cm} \scalebox{0.5}{\includegraphics[angle=0]{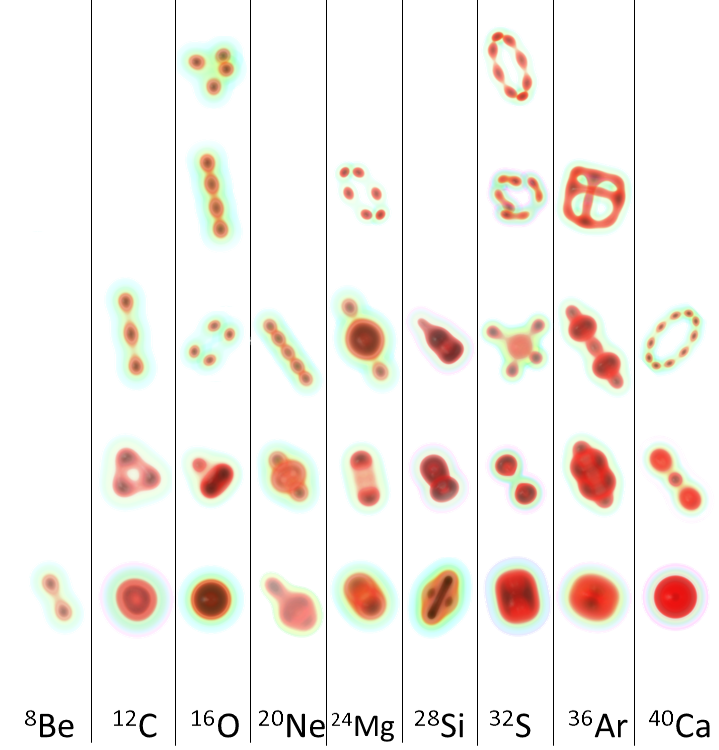}}
\caption{(Color online) Positive-parity projected density plots obtained for 
a number of excited configurations in $N=Z$ nuclei. 
For each nucleus the density in the bottom row 
corresponds to the equilibrium configuration. Other selected densities are 
displayed in order of increasing excitation energy.}
\label{fig:appb}
\end{figure} 
 \begin{figure*}[h!]
\centering

     \subfloat[$^8$Be $(K = 0)$ Mean-Field \label{fig:Be8_Eb2b3_mf}]{%
        \includegraphics[width=0.50\textwidth]{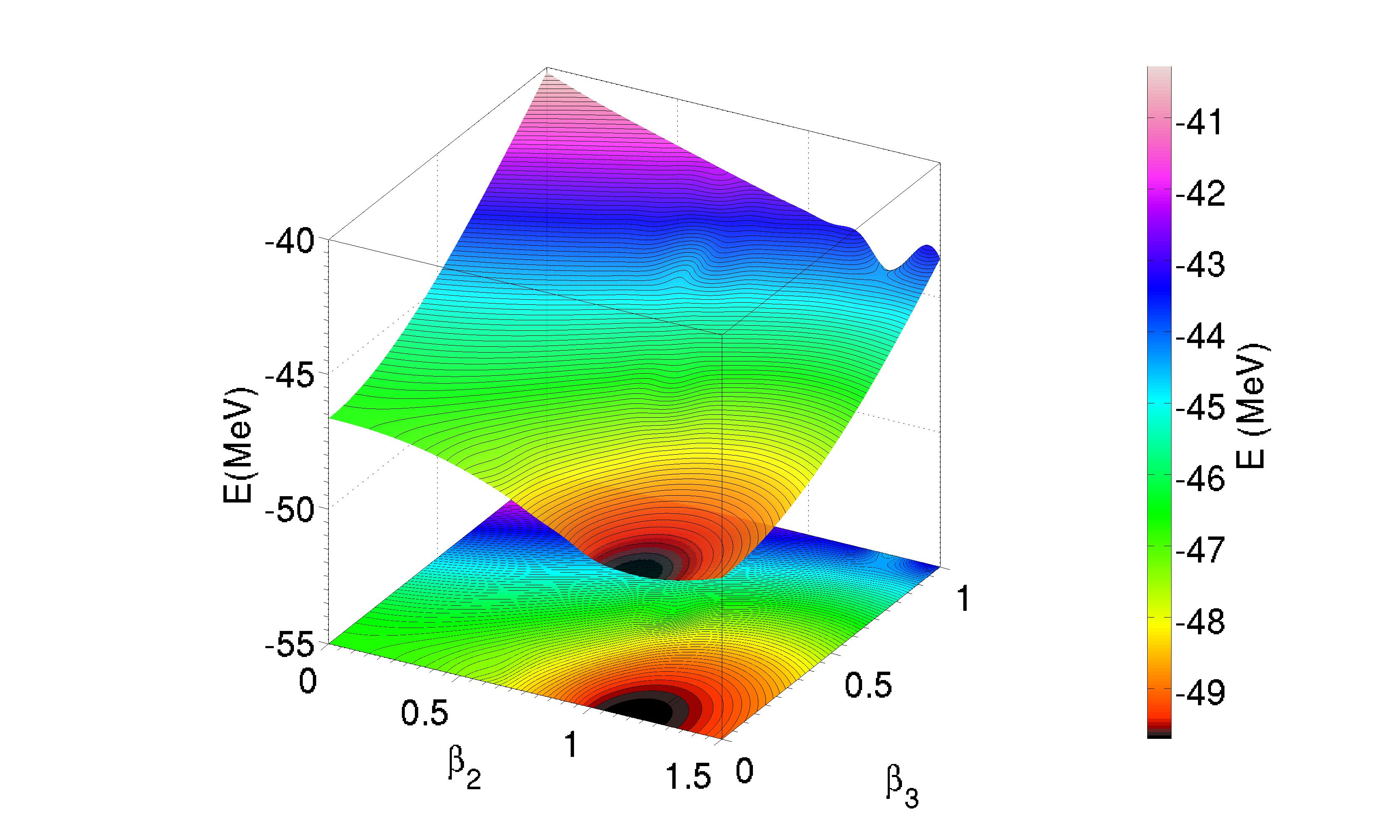}
      }\hspace{-4em}
      \subfloat[$^8$Be $(K^\pi = 0^+)$ PAV \label{fig:Be8_Eb2b3_projp}]{%
        \includegraphics[width=0.50\textwidth]{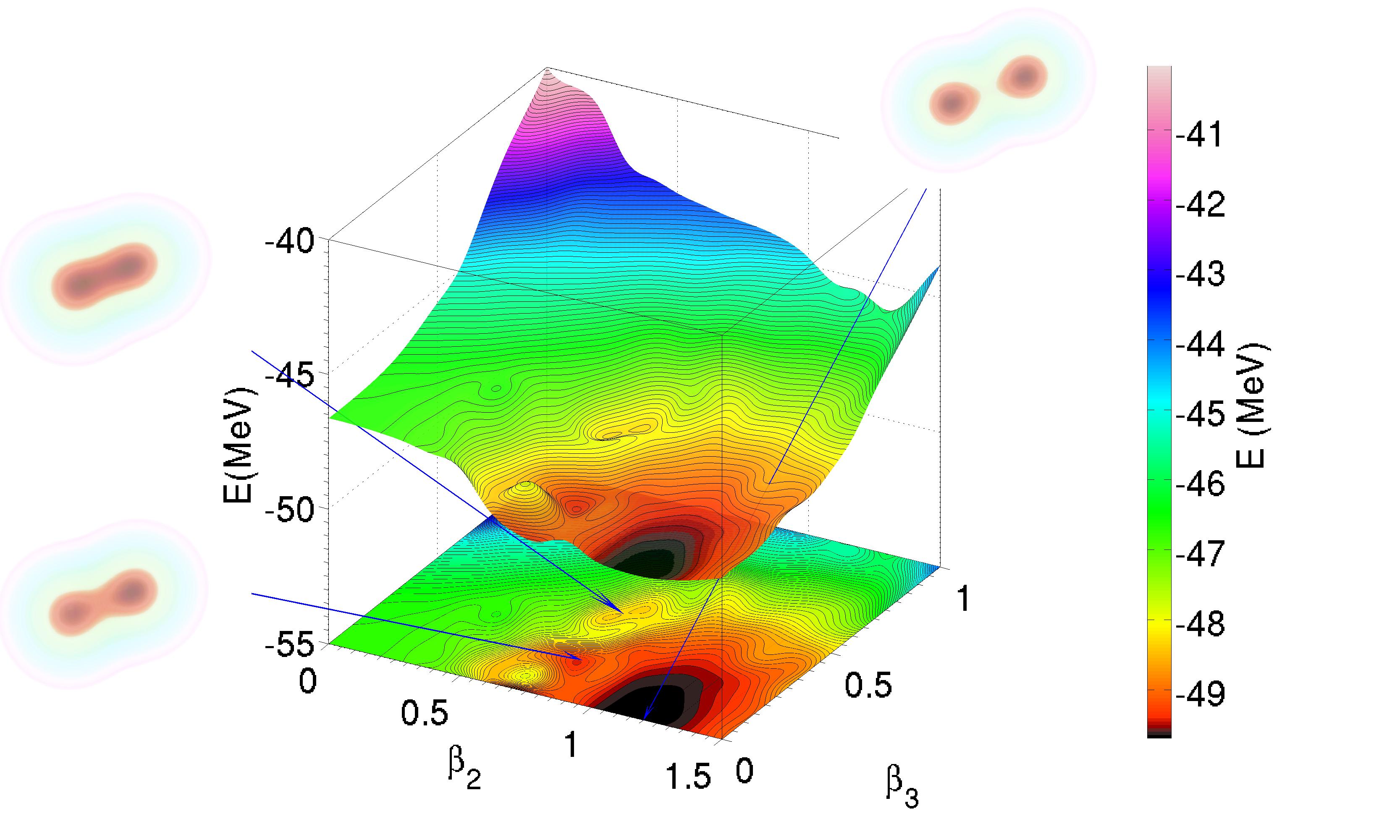}
      }


   \caption{(color online) Self-consistent energy surfaces of $^{8}$Be, 
calculated with DD-ME2 by imposing constraints on both the
axial quadrupole and octupole deformation parameters $\beta_2$ and
$\beta_3$ (left), and the corresponding positive parity-projected energy surfaces (right).}
    \label{fig:Beb2b3}
  \end{figure*}

\begin{figure*}[h!]
\centering

      \subfloat[$^{12}$C $(K = 0)$ Mean-Field \label{fig:C12_Eb2b3_mf}]{%
        \includegraphics[width=0.50\textwidth]{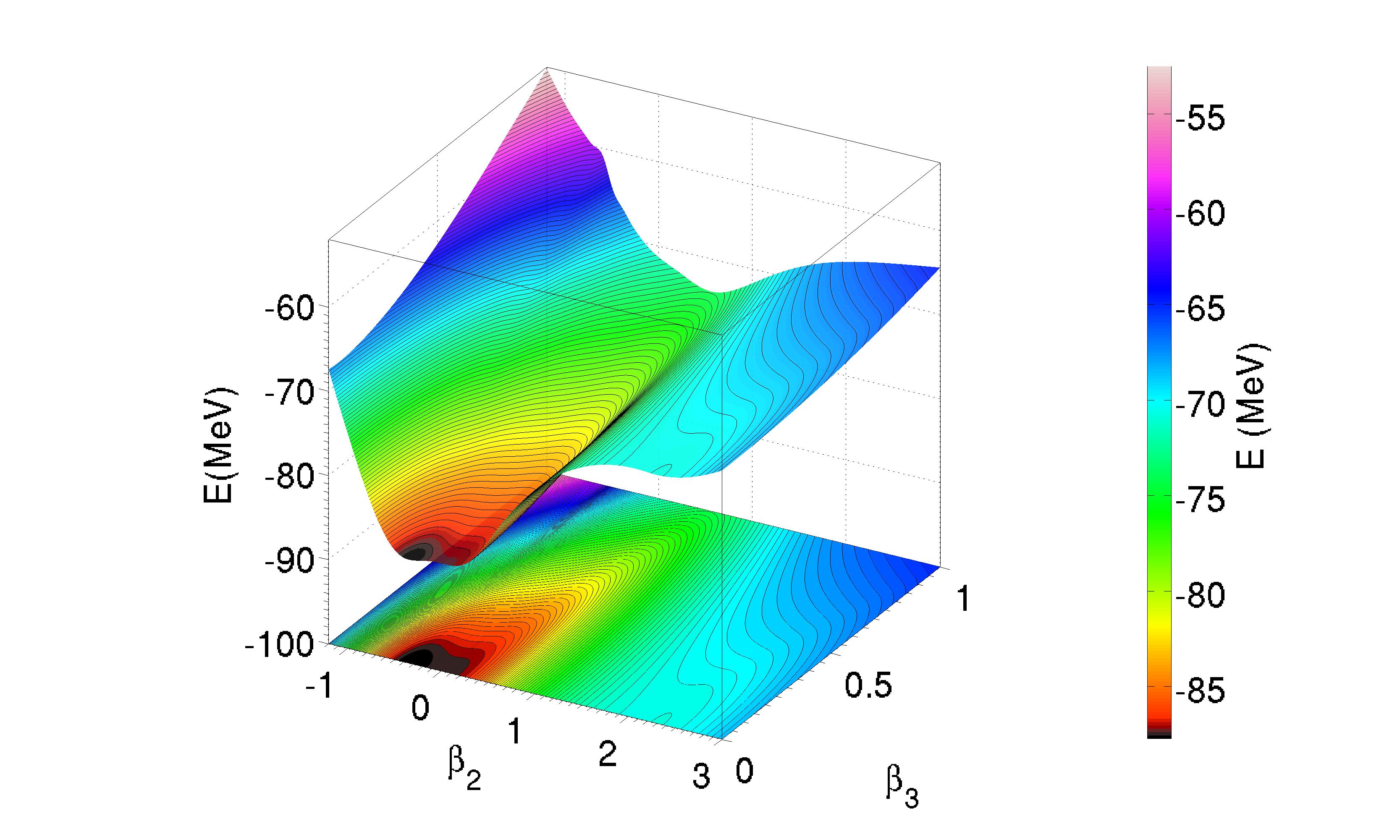}
      }\hspace{-4em}
      \subfloat[$^{12}$C $(K^\pi =0^+)$ PAV \label{fig:C12_Eb2b3_projp}]{%
        \includegraphics[width=0.50\textwidth]{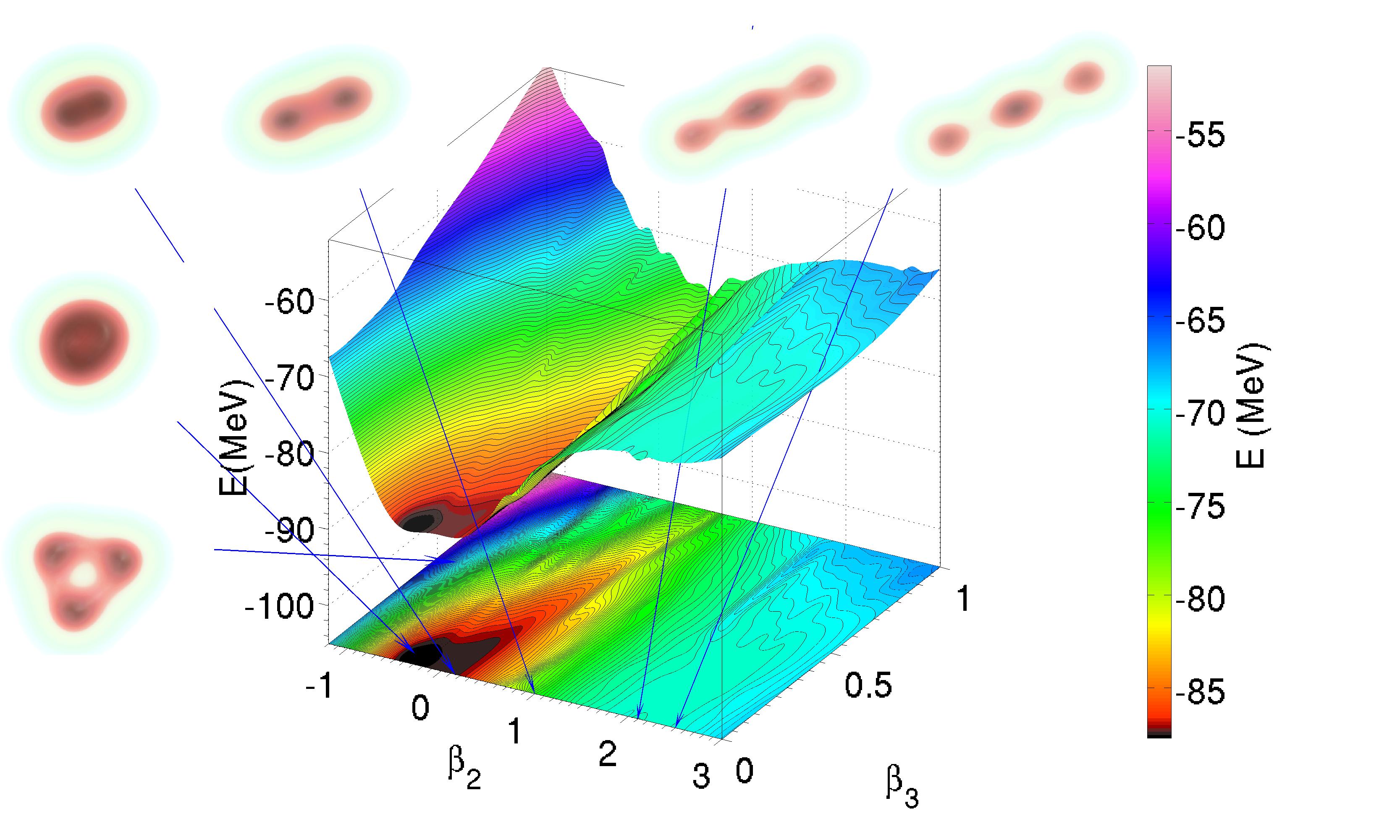}
      }
%
%
   \caption{(color online) Same as in the caption to Fig.~\ref{fig:Beb2b3} but for the 
   isotopes $^{12}$C.}
    \label{fig:Cb2b3}
  \end{figure*}

Let us illustrate this concept using the microscopic EDF framework
with the examples of axially deformed quadrupole shapes of $^{12}$C
and $^{20}$Ne. The self-consistent mean-field calculations with the 
relativistic functional DD-ME2 and a separable pairing interaction 
are performed using the implementation of the RHB model described in 
Ref.~\cite{Nik14}. The RHB equations are solved in the configurational 
space of harmonic oscillator wave functions with appropriate symmetry, 
whereas the densities are computed in coordinate space. The method 
can be applied to spherical, axially and non-axially deformed nuclei.
The map of the energy surface as a function of quadrupole deformation parameters is
obtained by solving the RHB equation with constraints on the axial and triaxial 
mass quadrupole moments of a given nucleus. The
method of quadratic constraints uses an unrestricted variation of the function
\begin{equation}
\langle \hat{H} \rangle + \sum_{\mu =0,2}{C_{2\mu} ( \langle \hat{Q}_{2\mu} \rangle
-q_{2\mu})^2},
\end{equation}
where $\langle \hat{H} \rangle$ is the total energy and $\langle \hat{Q}_{2\mu}\rangle$
denotes the expectation value of the mass quadrupole operators
\begin{equation}
\hat{Q}_{20} = 2z^2-x^2 - y^2 \quad \textnormal{and} \quad
\hat{Q}_{22} = x^2 - y^2 .
\end{equation}
$q_{2\mu}$ is the constrained value of the multipole moment and $C_{2\mu}$ the
corresponding stiffness constant~\cite{Rin80}.

By increasing the prolate quadrupole deformation in the 
axially symmetric 
self-consistent calculation with the constraint on the axial
quadrupole moment, $^{12}$C
and $^{20}$Ne display a series of cluster
configurations until eventually reaching the linear $\alpha$-chain
structure (Fig~\ref{fig:Emoychain}). To show the role of level
degeneracy, Fig~\ref{fig:Emoychain} displays the mean value of the
energy gap between consecutive occupied neutron single-particle levels
as a function of the deformation parameter $\beta_2$. The mean value
of the energy gap is defined as
\begin{equation}
\Delta\epsilon_n\equiv<\Delta\epsilon_i>
\end{equation}
where $\Delta\epsilon_i\equiv\epsilon_{i+1}-\epsilon_i$ is the energy
gap between two successive neutron single-particles levels. At deformations
for which the maximum mean value of the energy gap exceeds 5 MeV, we
only plot this value so that the scale of the vertical axis does not
become too large to display. A pronounced correlation between the
enhancement of energy gaps between the single-particle levels and
alpha-cluster formation can clearly be identified. Both for $^{12}$C
and $^{20}$Ne the density profiles show more pronounced localization of
$\alpha$ clusters at deformations at which the mean value of the
energy gap between consecutive levels exhibits a sharp increase.

\subsection{Quadrupole and octupole deformations 
and parity-projected energy surfaces}

Recent cranking SCMF calculations of high-spin rotating nuclei
produced interesting exotic cluster configurations such as, for
instance, in $^{16}$O and $^{40}$Ca \cite{ich11,ich12,fun10}. In the
present study cluster shapes occur as local minima at large
deformations on the ($\beta_2,\gamma,\beta_3,\beta_{32}$) energy
hypersurface. As an illustration, Figure \ref{fig:appb} displays a
sample of various cluster shapes in self-conjugate nuclei, obtained in
triaxial and reflection-asymmetric RHB calculations using the
functional DD-ME2. For each of the nuclei shown in Fig.~\ref{fig:appb}
densities that correspond to positive-parity projected intrinsic 
states are arranged in order of increasing energy. 
Most of them correspond to local minima on the deformation
energy surface, except for the ring states. The bottom row displays the lowest energy
(equilibrium) density distributions. This figure represents the
microscopic EDF-based analogue of the original Ikeda diagram, which
illustrates the coexistence of the nuclear mean-field and various
cluster structures that appear close to the (multi)
$\alpha$-separation threshold energies \cite{Ike68}. For instance,
already the equilibrium density of $^8$Be displays a two-$\alpha$
cluster configuration \cite{Bri66,Toh01,Wir00}. In the case of
$^{12}$C, the equilibrium self-consistent mean-field configuration
exhibits a slightly oblate triangular distribution of the three
$\alpha$ particles (i.e. the axial octupole moment does not vanish in 
the equilibrium configuration), which becomes much more pronounced in 
the excited configuration shown in the second row, in agreement with
very recent experimental results \cite{mar14}. At still higher
energies we find a linear chain configuration of the three $\alpha$
particles. $^{16}$O displays the very interesting 4$\alpha$ cluster
configuration with tetrahedral symmetry, a result very recently 
obtained using the constrained SCMF method \cite{gir13,ebr14}, the 
algebraic cluster model \cite{bij14}, and {\em ab initio} lattice
calculations using chiral nuclear effective field theory \cite{epe14}.
For heavier $Z=N$ nuclei, in Fig.~\ref{fig:appb} we include a variety of
exotic cluster configurations. For instance, as noted in the
original Ikeda description \cite{Ike68}, the lowest cluster
configuration of $^{20}$Ne corresponds to an $\alpha$ + $^{16}$O core
state. 

In the next section we will consider, in particular, the occurrence of
clusters in exotic Be and C isotopes. In the case of the N=Z nuclei,
the axial quadrupole and octupole nucleonic density distributions of
$^8$Be and $^{12}$C correspond to local minima on the energy surfaces
as functions of axial quadrupole and octupole deformations displayed
in Fig.~\ref{fig:Beb2b3} and Fig.~\ref{fig:Cb2b3}, respectively. The
self-consistent reflection-asymmetric axial energy surfaces are
calculated by imposing constraints on both the axial quadrupole and
octupole deformation parameters $\beta_2$ and $\beta_3$, respectively.
In addition with the constraint on the moment associated to the
octupole operator $\hat{Q}_{3}=r^{3}Y_{30}$, a constraint is also
imposed on the center of mass of the nucleus: $\left < r^{1}Y_{10}
\right > = 0$, to exclude the coupling to the spurious center of mass
motion. The 3D energy maps and their projections on the $\beta_2 -
\beta_3$ plane in the left part are obtained in SCMF calculations. The
corresponding positive parity-projected energy surfaces are shown in
the right part.  Positive ($\pi = +1$) and negative ($\pi = -1$)
parity-projected states are obtained by acting with the projector
$\hat{{\cal{P}}}^{\pi}$ on the intrinsic state: $| {\Phi}^{\pi}
(\beta_{2},\beta_{3}) \rangle = \hat{{\cal{P}}}^{\pi}|
{\Phi}(\beta_{2},\beta_{3}) \rangle$, where
\begin{equation}
\hat{{\cal{P}}}^{\pi} = \frac{1}{2} \left(1 + \pi \hat{\Pi} \right).
\end{equation}
The parity-projected energy surfaces are labelled with the 
deformation parameters of the intrinsic state and
calculated using the relation \cite{Rob07}:
\begin{widetext}
\begin{equation} 
E_{\pi} (\beta_{2},\beta_{3})=
\frac{
\langle {\Phi} (\beta_{2},\beta_{3}) | \hat{H} | {\Phi} (\beta_{2},\beta_{3}) \rangle
}
{
\langle {\Phi}(\beta_{2},\beta_{3}) |  {\Phi} (\beta_{2},\beta_{3}) \rangle
+
\pi \langle {\Phi} (\beta_{2},\beta_{3}) |  \hat{\Pi} | {\Phi} (\beta_{2},\beta_{3}) \rangle
}
+ \pi
\frac{\langle {\Phi} (\beta_{2},\beta_{3}) |
\hat{H}  \hat{\Pi} | {\Phi} (\beta_{2},\beta_{3}) \rangle
}
{
\langle {\Phi} (\beta_{2},\beta_{3}) |  {\Phi} (\beta_{2},\beta_{3}) \rangle
+
\pi \langle {\Phi} (\beta_{2},\beta_{3}) |  \hat{\Pi} | {\Phi} (\beta_{2},\beta_{3}) \rangle
}
\end{equation}
\end{widetext}
For the equilibrium deformations and few additional local minima the
nucleon density distributions in the reference frame defined
by the principal axes of the nucleus are shown in the insets. The
projected energy surface of $^8$Be displays a deep minimum at very 
large quadrupole deformation that corresponds to a two-$\alpha$
configuration in agreement with a number of previous studies
\cite{Bri66,Toh01,Wir00}. $^{12}$C offers the possibility to
investigate properties of three-center clusters. Linear chains of 
$\alpha$-particles are predicted at very large prolate quadrupole
deformations. A further possibility for three-center systems involves
the formation of triangular shapes characterized by a discrete
symmetry, and such structures are found in the region of oblate
deformations (cf. also Ref.~\cite{mar14}).


\section{Cluster structures in nuclei far from stability}

\begin{figure}[!htp]
\centering
\scalebox{0.30}{\includegraphics{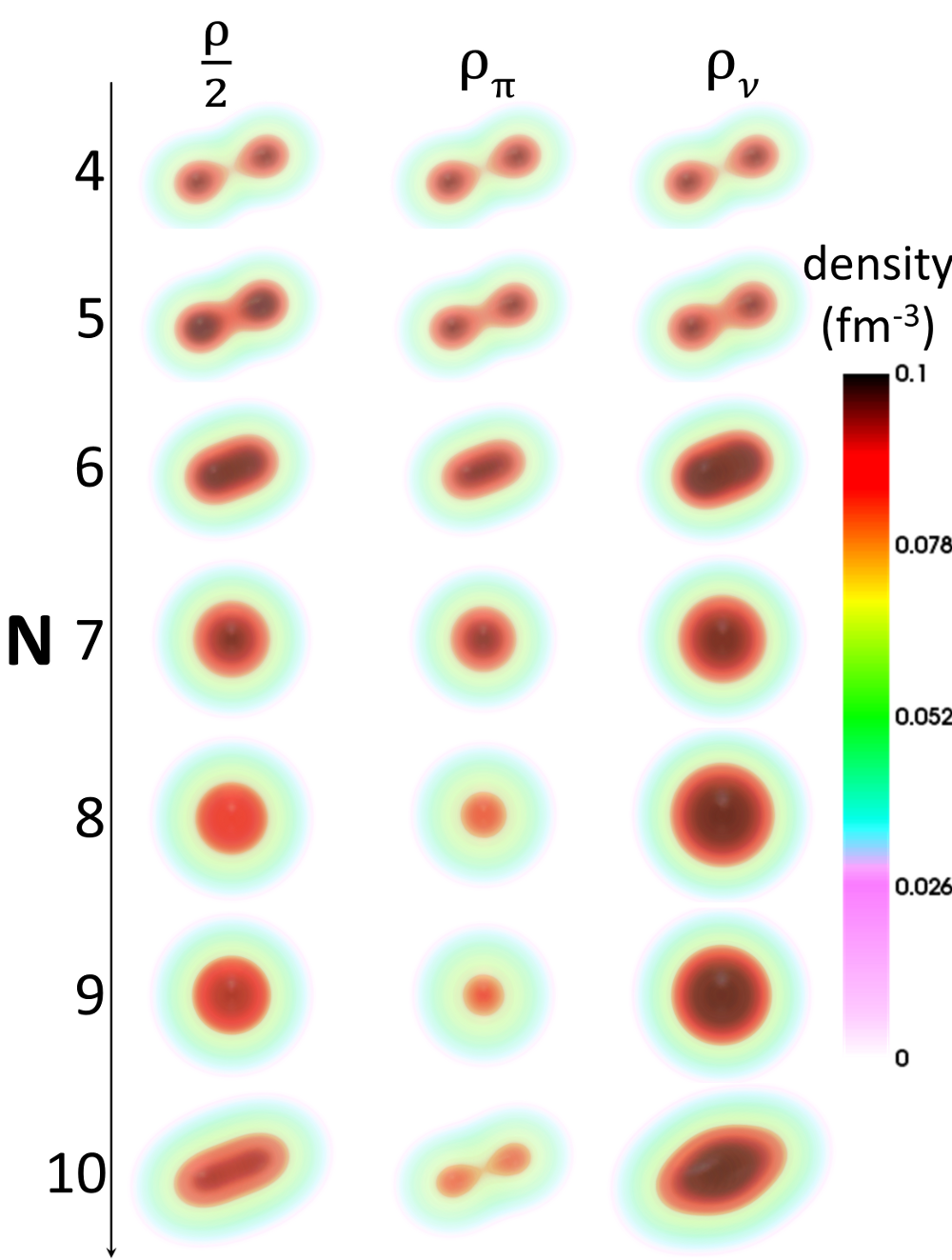}}
\caption{(Color online) Total, proton and neutron SCMF 
equilibrium intrinsic densities for beryllium isotopes,  
calculated using the RHB model with the functional DD-ME2.}
\label{fig:Bedensall}
\end{figure}

Low-energy structures in a number of relatively light neutron-rich
nuclei can be described by molecular bonding ($\pi$ or $\sigma$) of
$\alpha$-particles by the excess neutrons
\cite{oer96,oer01,4,5,Hor12,freer06,mil05,dec01,ito12}. Figure
\ref{fig:Bedensall} displays the total, proton and neutron axially 
symmetric intrinsic densities of Be isotopes in their equilibrium
configurations, calculated using the RHB model with the DD-ME2
functional. One clearly notices the two-$\alpha$ structure, except in
$^{10-13}$Be (the calculation for the odd-N isotopes is performed
using the equal filling approximation), which display nearly 
spherical shapes because of the N=8 shell closure. Even though 
recent experimental studies of charge radii and the corresponding Fermionic
Molecular Dynamics (FMD) calculations \cite{kri12} indicate a pronounced 
quenching of the N=8 shell in $^{12}$Be, a simple SCMF model based on 
a global functional that has not been specifically 
adjusted to this mass region, cannot produce such a structural change 
without additional adjustment of parameters and/or inclusion of 
correlations related to restoration of broken symmetries 
and configuration mixing. 

\begin{figure}[!htp]
\centering
\subfloat{%
      \includegraphics[width=0.60\textwidth]{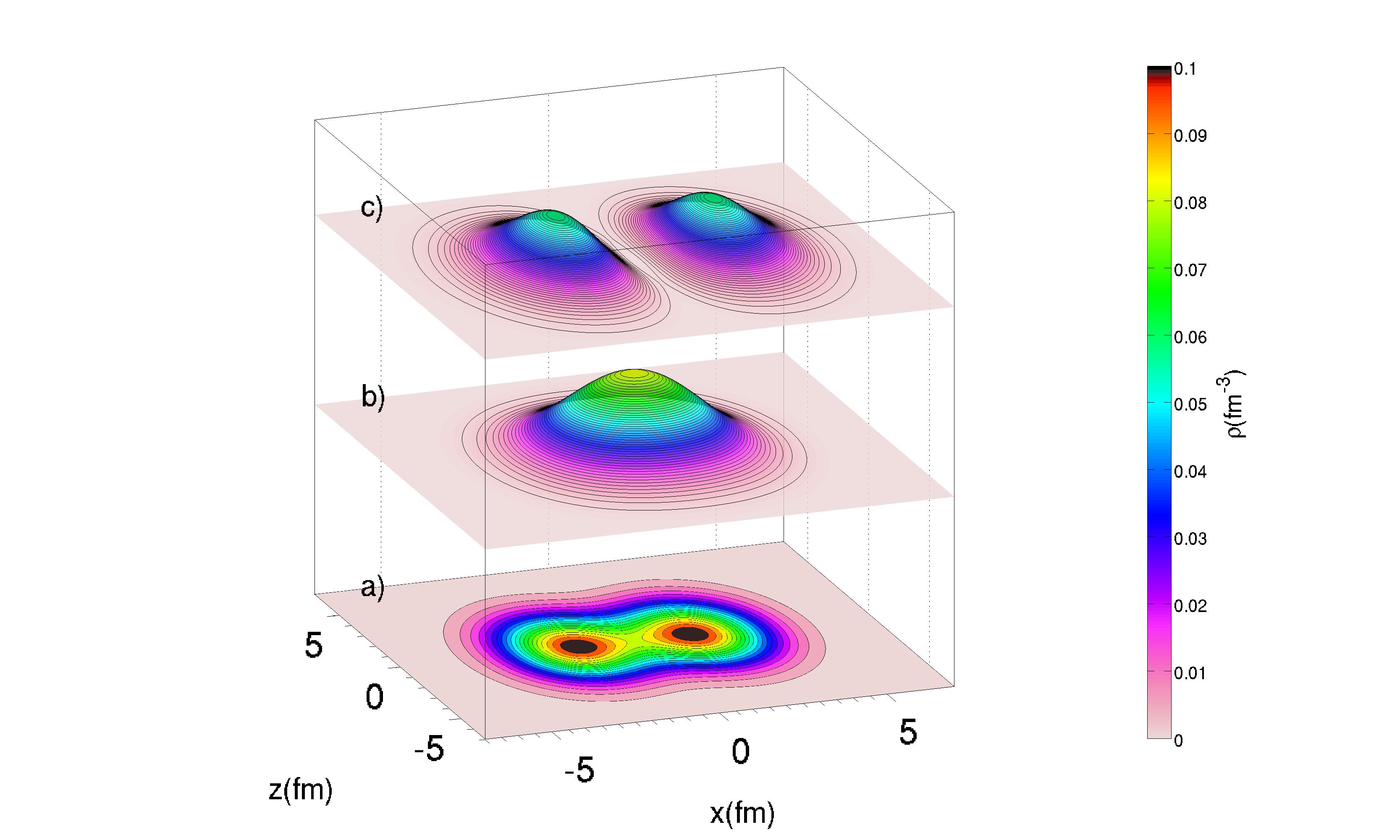}
    }
\caption{(Color online) Contour plot of the $^8$Be neutron density (a), and surface plots of the partial densities  
of each of the two occupied Nilsson states in the (Oxz) plane (b) and (c) ).}
\label{fig:Be08sp}
\end{figure}

To analyze the cluster content of Be isotopes, we investigate
the partial densities that correspond to occupied single-particle
states. Fig.~\ref{fig:Be08sp} displays the total neutron distribution 
of $^8$Be at equilibrium deformation, 
and details its decomposition into partial densities 
of each of the two occupied Nilsson states. A very similar 
picture is found for the proton density distributions. 
The partial densities 
provide a very clear picture of the formation of the two 
$\alpha$ clusters that appear in the total density distribution. 

\begin{figure}[!htp]
\centering
\scalebox{0.18}{\includegraphics{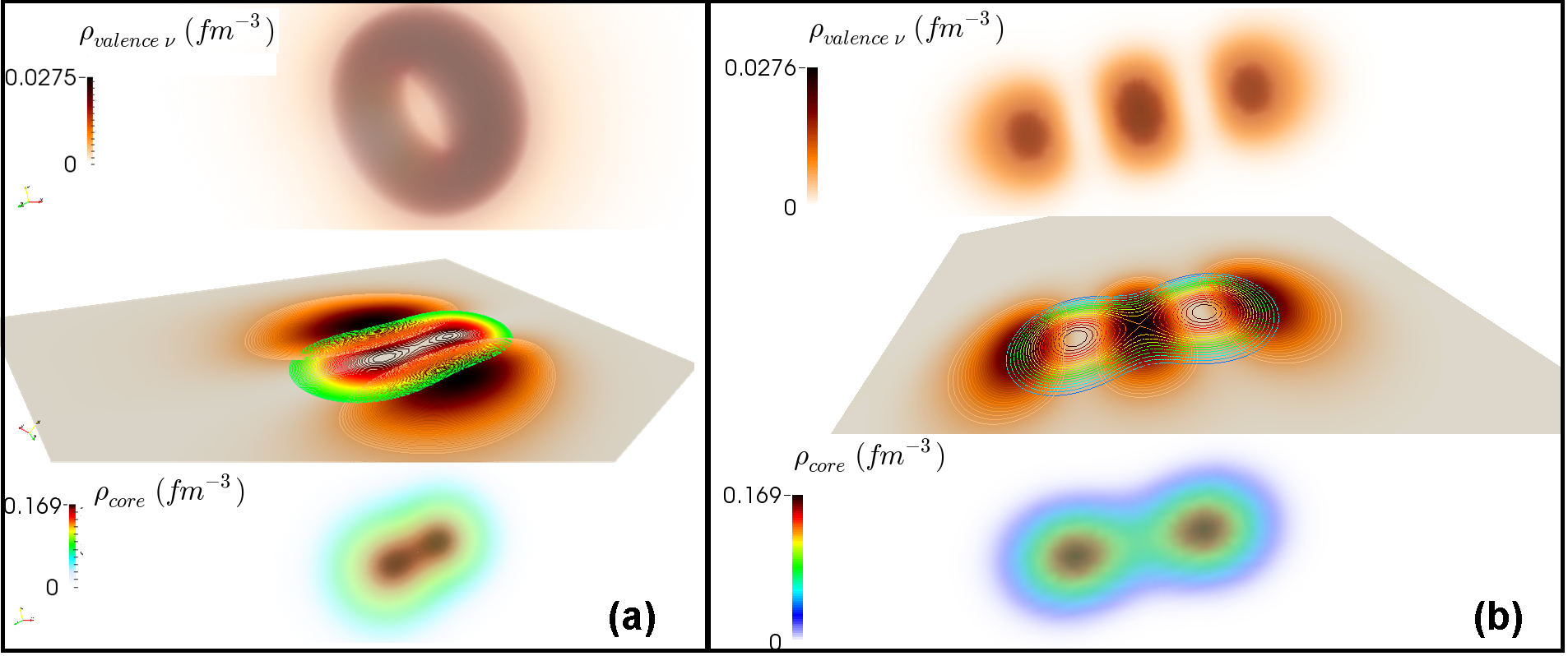}}
\caption{(Color online) Intrinsic 
densities of $^{10}$Be at equilibrium deformation (a), and for an excited configuration (b). 
From bottom to top: 
3D density of the $\alpha + \alpha$ core; contour plots of the core density and 
the density of the valence 
neutrons in the (Oxz) plane; 3D density of the valence neutrons.}
\label{fig:Be10all}
\end{figure}

\begin{figure}[!htp]
\centering
\scalebox{0.18}{\includegraphics{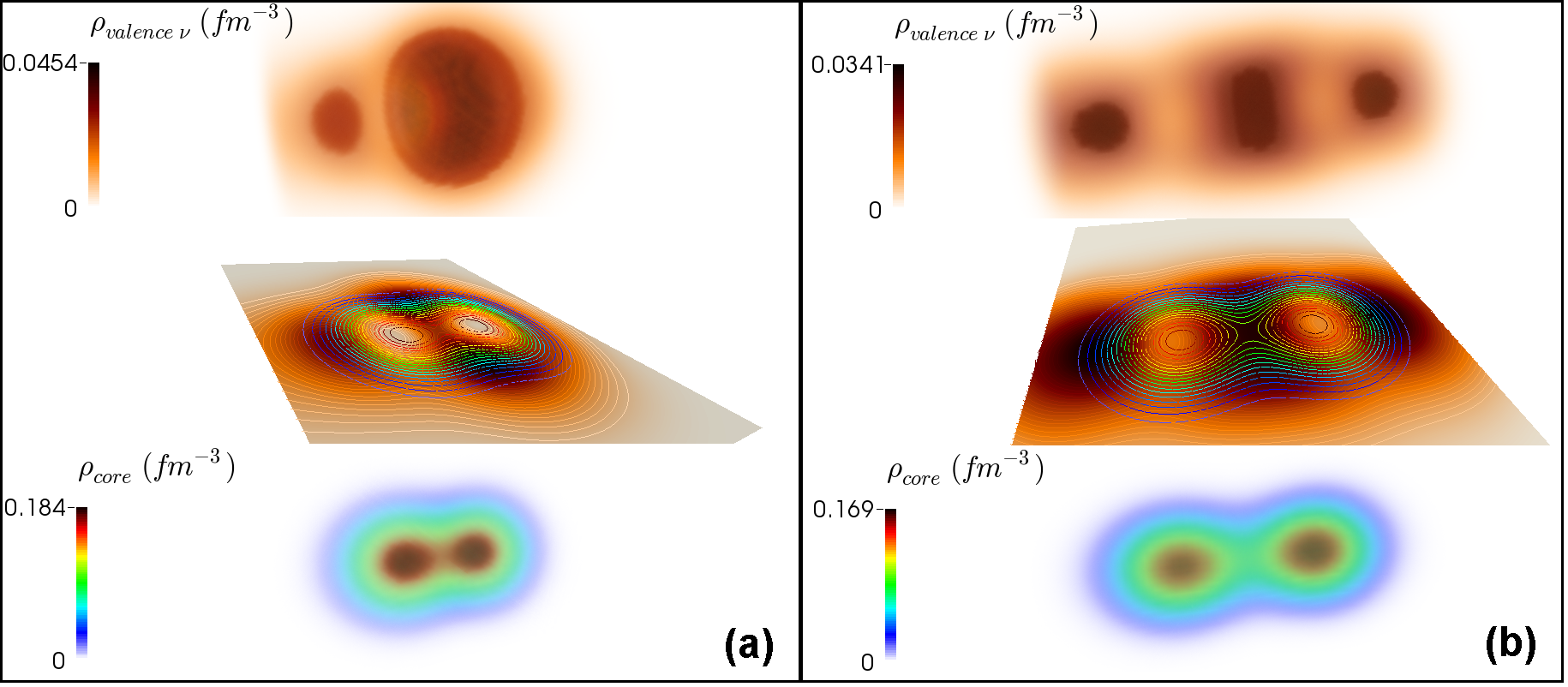}}
\caption{(Color online) Same as in the caption to Fig.~\ref{fig:Be10all} but for 
the equilibrium deformation and an excited configuration of $^{14}$Be.}
\label{fig:Be14all}
\end{figure}

In the case of the neutron-rich Be isotopes, decomposing the total
density into the $\alpha + \alpha$ structure and the density of 
the additional valence neutrons, a picture of nuclear molecular 
states emerges. A negative-parity orbital perpendicular to the $\alpha + \alpha$ 
direction is called a $\pi$-orbital, and a positive-parity orbital  parallel to the 
$\alpha + \alpha$  direction is called a $\sigma$-orbital 
(cf. Fig. 7 of Ref.~\cite{yos12}).
As an example here we consider 
$^{10}$Be and $^{14}$Be. The valence neutrons
stabilize the two-center cluster structure of the $\alpha + \alpha$ core
with $\pi$-like and $\sigma$-like molecular bonds (Figs.~\ref{fig:Be10all} and \ref{fig:Be14all}). 
The results obtained in the present calculation 
are consistent with predictions of the Antisymmetrized Molecular
Dynamics model (cf. Ref.~\cite{yos12} and references cited therein), that is, 
the valence neutrons form a $\pi$-bond in the equilibrium state and a $\sigma$-bond in the excited
state shown in Fig.~\ref{fig:Be10all}. In the case of the more neutron-rich nucleus $^{14}$Be, 
as shown in Fig.~\ref{fig:Be14all}, already in the equilibrium state the valence neutrons
form both $\pi$ and $\sigma$ bonds, similar to the results reported in Ref. \cite{yos12}. 

\begin{figure}[!htp]
\centering
        \includegraphics[width=0.45\textwidth]{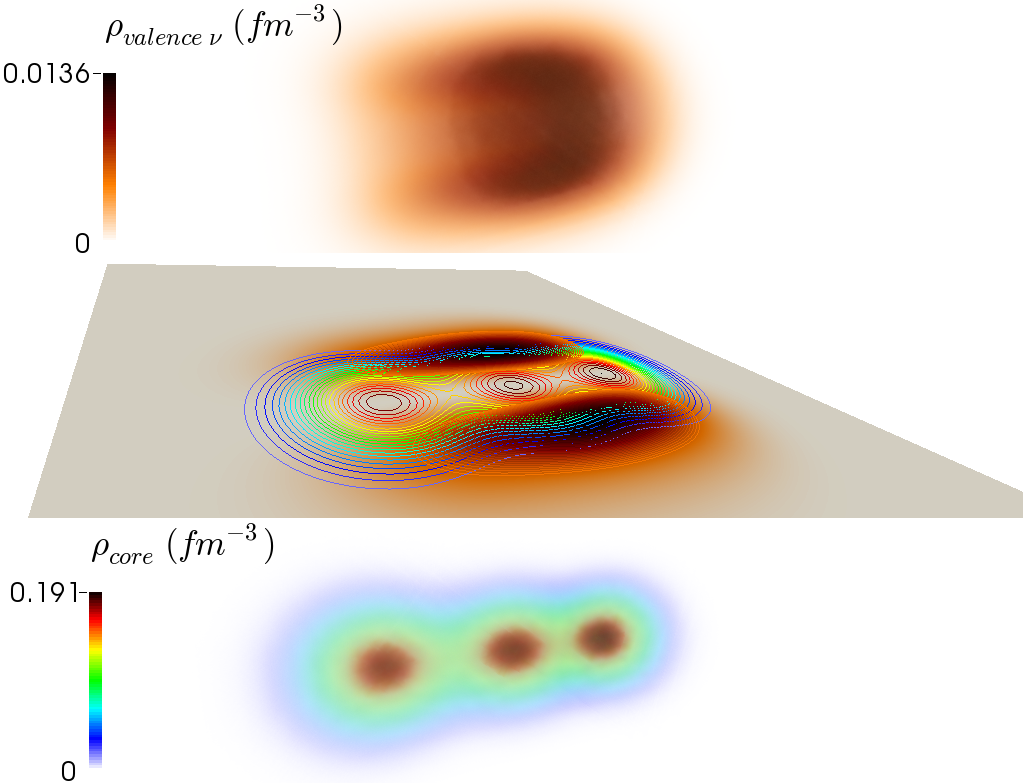}
\caption{(Color online) Nucleonic densities for an excited configuration of $^{14}$C. From bottom to top: 
3D density of the $\alpha + \alpha + \alpha$ core; contour plots of the core density and 
the density of the valence 
neutrons in the (Oxz) plane; 3D density of the valence neutrons.}
\label{fig:C14chain}
\end{figure}

\begin{figure}[!htp]
\centering
\scalebox{0.25}{\includegraphics{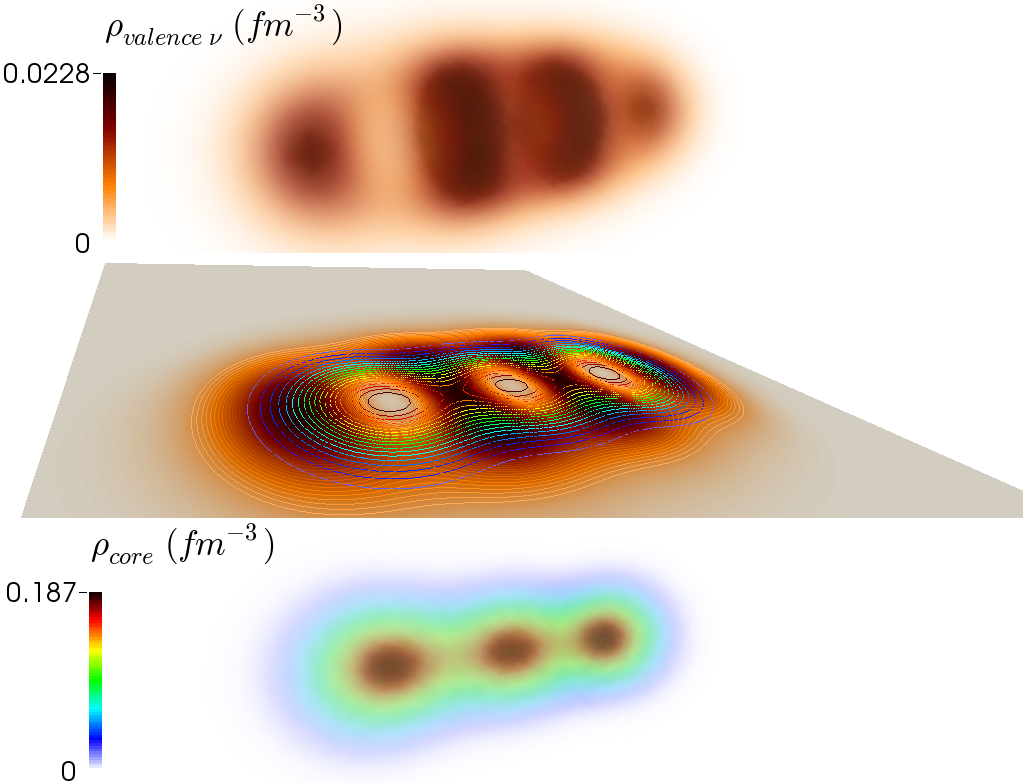}}
\caption{(Color online) Same as in the caption to Fig.~\ref{fig:C14chain} but 
for an excited configuration of $^{16}$C.}
\label{fig:C16chain}
\end{figure}

An interesting topic is the occurrence of clusters in excited states of 
neutron-rich carbon isotopes \cite{5}. In particular, the molecular-orbital structure 
in neutron-rich C isotopes was investigated using a microscopic 
molecular-orbit (MO) $\alpha + \alpha + \alpha + n + n +\ldots$ model 
\cite{ita01}, and it was shown that valence neutrons which occupy the $\pi$-orbit
increase the binding energy and stabilize the linear chain of 3 $\alpha$ against the 
breathing-like breakup. However, when considering $^{12}$C, $^{14}$C, and $^{16}$C, 
it was found that the linear-chain structure of 
$^{16}$C $((3/2^-_\pi)^2 (1/2^-_\sigma)^2)$ is the only one to be simultaneously 
stable against the breathing-like breakup and the bending-like breakup. 
Figs.~\ref{fig:C14chain} and
\ref{fig:C16chain} display the excess-neutron molecular orbits 
in excited configurations of $^{14}$C and $^{16}$C, calculated using the 
present EDF-based self-consistent microscopic approach. 
The decomposition of the density of an excited configuration of
$^{14}$C  in terms of the 3 $\alpha$ core
and the two valence neutrons is shown in Fig~\ref{fig:C14chain}. We note
that in this case correlations between the valence
neutrons tend to favor a reflection asymmetric chain configuration.
Accordingly, the intrinsic reflection-asymmetric chain configuration
$\alpha-2n-\alpha-\alpha$, with the two valence neutrons forming a
$\pi$-bond between the two $\alpha$, is found at lower energy with respect
to the reflection symmetric chain $\alpha-n-\alpha-n-\alpha$. A
reflection-symmetric configuration with four valence neutrons, 
that is,  $\alpha-2n-\alpha-2n-\alpha$ is favored in $^{16}$C, as 
shown Fig~\ref{fig:C16chain} and similar to the results obtained with 
the AMD approach \cite{5}. 

\begin{figure}[!htp]
\centering
\scalebox{0.20}{\includegraphics{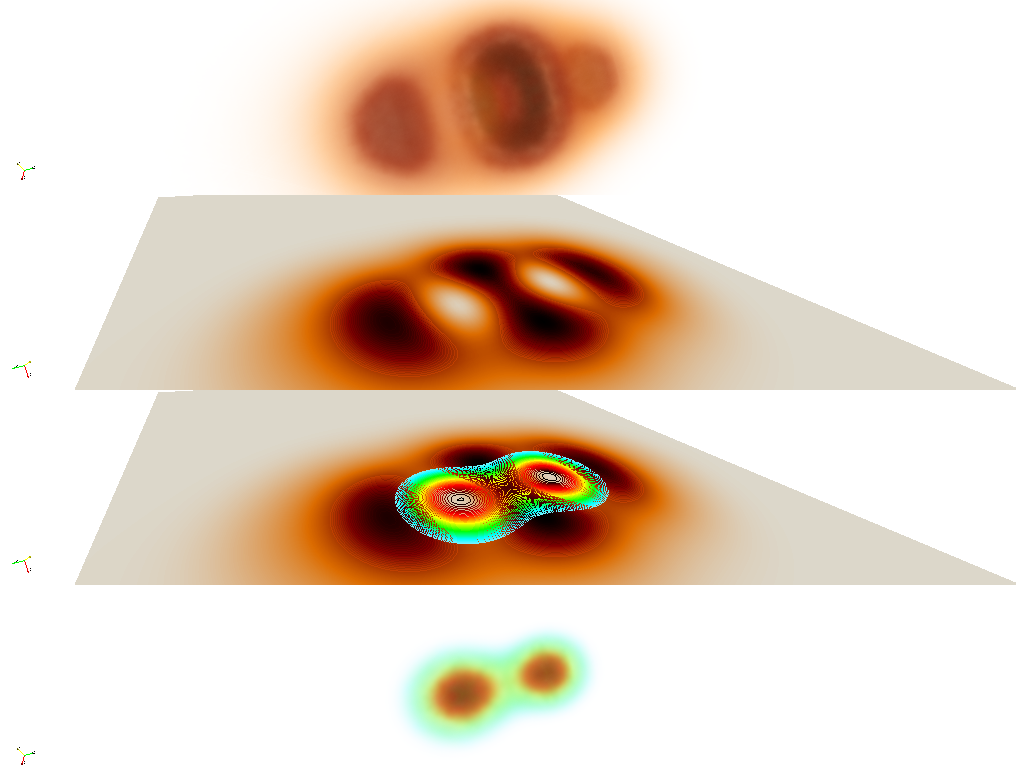}}
\caption{(Color online) Nucleonic densities for an excited configuration of $^{10}$C. From bottom to top: 
3D density of the $\alpha + \alpha $ core; contour plots of the core density and 
the density of the valence 
protons in the (Oxz) plane; 3D density of the valence protons.}
\label{fig:C10chain}
\end{figure}

On the proton-rich side a particularly interesting case is $^{10}$C, 
which may be described as an $\alpha + \alpha + p + p$ structure. 
The unique feature of this system is that the removal of any one of the four 
constituents results in an unbound three-body system. It can, therefore, 
be considered as a super-Borromean or fourth-order 
Brunnian nuclear system \cite{char07,cur08}. As
the mirror nucleus of $^{10}$Be, $^{10}$C is expected to display a
covalent 2-center chain configuration with a pair of protons as the
covalent bond. Fig.~\ref{fig:C10chain} illustrates the results of our 
EDF-based self-consistent calculation for an excited configuration of $^{10}$C, 
for which the two valence protons provide the molecular bonding for the 
$\alpha + \alpha $ core.

\section{Conclusion}
The formation of cluster states in nuclei has been investigated employing a theoretical framework 
based on nuclear EDF. By performing deformation constrained self-consistent 
Hartree-Fock-Bogoliubov (HFB)
calculations with Skyrme and Gogny functionals, and relativistic Hartree-Bogoliubov (RHB) calculations 
with the functional DD-ME2, it has been shown that a deeper  
mean-field confining potential leads to a more pronounced localization of the single nucleon wave functions 
and enhances the probability of formation cluster structures in excited states. In particular, since the relativistic 
functional DD-ME2 produces the deepest potential among the considered functionals, we have 
used DD-ME2 in a series of axially-symmetric quadrupole and octupole constrained RHB calculations 
of relatively light $N=Z$, as well as neutron-rich nuclei. The role of deformation and degeneracy of 
single-nucleon states in the formation of clusters has been analyzed in detail, and a number of interesting 
cluster structures have been predicted in excited configurations that correspond to 
local minima on the parity-projected energy maps as functions of the quadrupole and octupole deformation 
parameters. 

A particularly interesting topic is the occurrence of cluster configurations in neutron-rich nuclei. 
We have shown that in neutron-rich Be and C nuclei cluster states occur as a result of 
molecular bond ($\pi$ or $\sigma$) of $\alpha$-particles by the excess neutrons, and 
also that proton covalent bond can occur in $^{10}$C. 

Results obtained in this study demonstrate the feasibility of using
nuclear EDF to explore the occurrence and
evolution of $\alpha$-cluster structures in relatively light $N=Z$ and
neutron-rich nuclei. When compared to dedicated cluster models, this 
framework allows for a microscopic description of the coexistence of
cluster states and mean-field-type states at low energies. The SCMF
approach does not assume any constraint on the nucleonic wave
function nor the existence of nucleon
cluster structures, rather energy density functionals implicitly
include many-body correlations that enable the formation of cluster
states starting from microscopic single-nucleon degrees of freedom.
For a quantitative description of cluster states, however, EDF-based
structure models have to be developed that go beyond the static
mean-field approximation, and include collective correlations related 
to the restoration of symmetries broken at the mean-field level, and
to fluctuations of collective variables.  These models can then be
employed in analyses of cluster phenomena related to shell evolution 
and shape transitions, including detailed predictions of excitation
spectra and electromagnetic transition rates. 


\section*{Acknowledgement}

This work was supported by the Institut Universitaire de France. The
authors thank Peter Schuck and Matko Milin for reading the manuscript, many valuable 
discussions and suggestions.
\appendix
\section{The localization and quantality parameters}

The localization parameter is used to characterize clusters as hybrid
states between the crystal and quantum liquid phases in nuclei \cite{nat,ebr13,ebr14},
whereas the quantality parameter describes the quantum liquid vs.
crystal behavior of homogeneous nucleonic matter \cite{mot96} and is
defined by the relation:
\begin{equation}
\label{eq:lam}
\Lambda\hat{=} \frac{\hbar^2}{m r_0^2V'_0}  \;
\end{equation}
where r$_0$ is the typical inter-nucleon distance and V'$_0$ the
characteristic magnitude of the inter-particle interaction (V'$_0\simeq$100
MeV in the case of the nucleon-nucleon interaction). As discussed by
Mottelson, the quantum liquid phase is obtained for $\Lambda > 0.1$, 
whereas the crystal phase is characterized by values of $\Lambda < 0.1$. 
Nuclei, of course, are in the quantum liquid phase. However, 
the quantality parameter Eq. (\ref{eq:lam}) depends on
the nucleon-nucleon interaction only, and does not take into account 
the finite size effects at work in nuclei. Hence the localization
parameter is defined as \cite{nat,ebr13,ebr14}
\begin{equation}\label{eq:alpha}
\alpha\hat{=}\frac{b}{r_0}=\frac{\sqrt{\hbar}A^{1/6}}{(2mV_0r_0^2)^{1/4}}
\end{equation}
where b is the typical dispersion of the single-nucleon wave function, and
V$_0$ is the depth of the confining potential (V$_0\simeq$75 MeV
in the case of the nuclear mean-field \cite{nat}). One can therefore use $\alpha$ to 
analyze the evolution of nuclear configurations with respect to the number of 
constituents A \cite{ebr13,ebr14} and, in particular, systems where finite-size
effects are relevant (A $< 10^3$). The crystal, cluster and
liquid phases then correspond to $\alpha < 1$, $\alpha\sim 1$, and $\alpha > 1$,
respectively.

In order to relate the quantality and localization parameters, we need 
to link the depth of the
mean-field potential $V_0$ to the magnitude of the nucleon-nucleon
interaction $V'_0$. Considering a short range n-n interaction
$V_2(\vec{r},\vec{r'})$ that can qualitatively be described by a
hard core for r$<$r$_0$ and an attractive part of magnitude $-V'_0$
in the region between r$_0$ and r$_0$+a (Fig.~\ref{fig:pot}):

\begin{figure}[tb]
\begin{center}
\scalebox{0.35}{\includegraphics{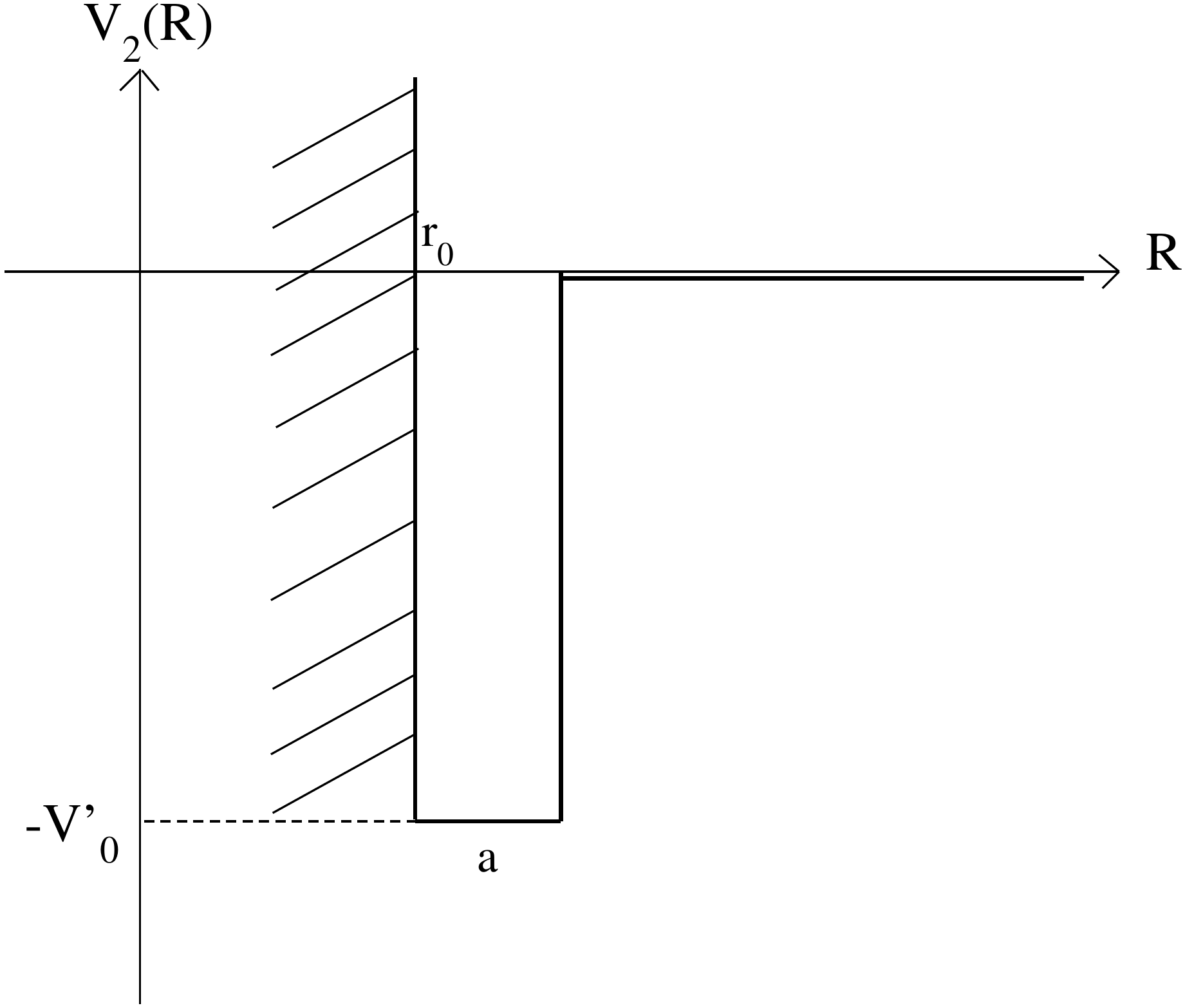}}
\caption{Simple approximation of the n-n potential used to 
derive the relation between the localization and quantality 
parameters Eq.~(\ref{eq:lq}).}
\label{fig:pot}
\end{center}
\end{figure}

\begin{equation}
V_2(\vec{r},\vec{r'})=V_2(R)=-V'_0
\label{eq:v2}
\end{equation}
for R between r$_0$ and r$_0$+a, with $R\equiv |\vec{r}-\vec{r'}|$.

The n-n interaction can also be approximated by
\begin{equation}
V_2(\vec{r},\vec{r'})=-V''_0\delta(\vec{r}-\vec{r'}-\vec{r_0})=-V''_0\delta(\vec{R}-\vec{r_0})\;.
\label{eq:v2a}
\end{equation}
This can be justified by the short range-approximation of the
nucleon-nucleon interaction, and such a zero-range 
approximation is successfully used, for instance, in Skyrme functionals. To be
compatible with Eq. (\ref{eq:v2}), a$<$r$_0$.

The confining potential V(r) is, to a good approximation, the mean
value of the n-n interaction over the nucleonic density \cite{Rin80}:
\begin{equation}
V(\vec{r}) \simeq \int V_2(\vec{r},\vec{r'})\rho(\vec{r'})
d\vec{r'}\\
=-V''_0\rho(\vec{r}-\vec{r_0})
\label{eq:v}
\end{equation}
Eq. (\ref{eq:v}) expresses the fact that in a saturated system characterized by a 
short-range interaction, the mean-field potential displays a spatial dependence 
that corresponds to the shape of the density.
From Eq. (\ref{eq:v}) the depth of the mean-field potential is:
\begin{equation}
V_0 \approx V''_0\rho_0 
\label{eq:vv}
\end{equation}
where $\rho_0=3/(4\pi r_0^3)$.
Moreover, Eq. (\ref{eq:v2}) and (\ref{eq:v2a}) yield 
\begin{equation}
\int V_2(R)d\vec{R}=-V''_0=-4\pi V'_0 \int_{r_0}^{r_0+a} R^2 dR \;,
\end{equation} 
and thus
\begin{equation}
V''_0=\frac{4}{3}\pi V'_0 \left[ (r_0+a)^3-r_0^3\right] \;.
\label{eq:vs}
\end{equation}
Inserting Eq. (\ref{eq:vs}) in (\ref{eq:vv}), one finally obtains
\begin{equation}
V_0=\gamma V'_0
\label{eq:gama}
\end{equation}
with 
\begin{equation}
\gamma \equiv \left[\left(1+\frac{a}{r_0}\right)^3 -1 \right ]\;.
\label{eq:gam}
\end{equation}

Therefore, the relation between the depth of the mean-field
potential V$_0$ and the magnitude of the n-n interaction V'$_0$ is
linear and only depends, in this simple approximation, on the ratio a/r$_0$,
that is, the width of the attractive part of the n-n interaction
over the equilibrium distance between the nucleons. In finite nuclei, for
typical values of r$_0$ and $a$ one gets $\gamma \simeq$3/4. This is
in agreement with the empirical values V$_0$=75 MeV and V'$_0$=100 MeV
\cite{mot96,jam69}. Inserting now Eq. (\ref{eq:lam}) into Eq. (\ref{eq:alpha}), and 
using Eq. (\ref{eq:gama}) with $\gamma$= 3/4, one finds the relation between the localization 
and quantality parameters
\begin{equation} 
\alpha\simeq A^{1/6}\Lambda^{1/4}\;.
\label{eq:lq}
\end{equation}

\end{document}